\documentclass[12pt]{iopart}
\usepackage{cite}
\usepackage[pdftex]{graphicx}
\usepackage{float}
\usepackage[caption = false]{subfig}
\usepackage{hyperref}
\usepackage{iopams}

\graphicspath{{.}{./simulation/}}

\begin{document}
\title{Dim and bright void regimes in capacitively-coupled RF complex plasmas}
\author{A Pikalev$^1$, I Semenov$^2$, M Pustylnik$^1$, C R\"ath$^1$ and H Thomas$^1$}
\address{$^1$ Institut f\"ur Materialphysik im Weltraum, Deutsches Zentrum f\"ur Luft- und Raumfahrt e.\,V. (DLR), 82234 We{\ss}ling, Germany}
\address{$^2$  Leibniz Institute for Plasma Science and Technology, 17489 Greifswald, Germany}
\ead{Aleksandr.Pikalev@dlr.de}

\begin{abstract}

We demonstrate experimentally that the void in capacitively-coupled RF complex plasmas can exist in two qualitative different regimes.
The ``bright'' void is characterized by bright plasma emission associated with the void,
whereas the ``dim'' void possesses no detectable emission feature.
The transition from the dim to the bright regime occurs with an increase of the discharge power and has a discontinuous character.
The discontinuity is manifested by a kink in the void size power dependencies.
We reproduce the bright void (mechanically stabilized due to the balance of ion drag and electrostatic forces) by a simplified time-averaged 1D fluid model.
To reproduce the dim void, we artificially include the radial ion diffusion into the continuity equation for ions,
which allows to mechanically stabilize the void boundary due to very weak electrostatic forces. 
The electric field at the void boundary occurs to be so small that it, in accordance with the experimental observation, causes no void-related emission feature.

\end{abstract}


\maketitle
\ioptwocol

\section{Introduction}

Complex or dusty plasma is a medium containing ionized gas and micron-sized solid particles.
It is used in basic research as a model system for particle-resolved studies of generic condensed matter phenomena 
\cite{fortov2005complex, Morfill-RevModPhys2009, Ivlev2012}.
For such studies, homogenous microparticle suspensions are desirable. 
Therefore, many of the complex plasma experiments are performed in microgravity conditions \cite{Nefedov-NewJPhys2003, Thomas-NewJPhys2008, Pustylnik-RevSciInstrum2016}.
However, microgravity does not guarantee homogeneous distribution of the microparticle component. 
The appearance of a void, i.~e. a microparticle-free area, is one of most common disturbances of the complex plasma homogeneity under microgravity conditions 
\cite{Land-NewJPhys2008, Schmidt-AIPPhysPlas2011}.
The void formation and growth also determine the nanoparticle generation cycle in plasma reactors \cite{Stefanovic-PlasmaSrcSciTech2017}.
Although the void formation has been studied for at least two decades, understanding of its behaviour is still incomplete.

One of the difficulties in studying the void (as well as complex plasmas in general) is that the microparticles immersed into the gas discharge plasma change its properties. 
It was noticed already at the beginning of complex plasma research 
\cite{Bouchoule-PlasmaSrcSciTech1993, Bouchoule-PlasmaSrcSciTech1994, Tachibana-PlasmaSrcSciTech1994, Fridman-JApplPhys1996}. 
This influence  was confirmed and detailized in recent works \cite{Stefanovic-PlasmaSrcSciTech2017, Mitic-NewJPhys2009, Killer-AIPPhysPlas2013, Pustylnik-PhysRevE2017}.

The results of void closure experiments in \cite{Lipaev-PhysRevLett2007} were interpreted in terms of a single-particle approach, 
i.e., the force balance was treated for a single microparticle immersed in the simulated field of microparticle-free plasma parameters. 
According to \cite{Lipaev-PhysRevLett2007}, the void forms due to the balance of electrostatic and ion drag forces.
This idea, even nowadays, remains the baseline in the understanding of the physics of the void. 

In theoretical work \cite{Annaratone-PhysRevE2002}, a double layer formation and time averaged flows of the electrons and ions
in the vicinity of the void boundary were investigated.
In works \cite{Goree-PhysRevE1999, Tsytovich-PhysRevE2001, Tsytovich-PhysRevE2004, Vladimirov-AIPPhysPlas2005},
a theory of a spherical void in an infinite microparticle suspension was developed. 
No RF electron dynamics was considered.

To get better insight into microparticle-plasma interaction, the entire discharge was simulated in
\cite{Gozadinos-NewJPhys2003, Akdim-PhysRevE2003, Land-NewJPhys2007, Goedheer-PlasmaPhysControlFusion2008, Land-NewJPhys2008, Goedheer-JPhysD2009, Goedheer-ContribPlasPhys2009,
 Killer-AIPPhysPlas2013, Pustylnik-PhysRevE2017}.
In \cite{Gozadinos-NewJPhys2003}, quite a comprehensive 1D fluid model was developed. 
It included two primary physical mechanisms of plasma-microparticle interaction, 
namely the absorption of plasma electrons and ions on the microparticles 
and the contribution of the microparticles to plasma quasineutrality. 
This model reproduced the appearance of the void with reducing the pressure and a peak of microparticle density at the void boundary
often observed in the experiments.

2D fluid models were presented in \cite{Akdim-PhysRevE2003, Land-NewJPhys2007, Goedheer-PlasmaPhysControlFusion2008, Land-NewJPhys2008}.
These models demonstrated the increase of the electron energy in the presence of the microparticles.
If the void existed in the model, the electron density inside the void was much higher than inside the microparticle suspension.
The simulated void size was compared in \cite{Goedheer-PlasmaPhysControlFusion2008, Land-NewJPhys2008} with the experiment \cite{Lipaev-PhysRevLett2007}.
Although the model qualitatively reproduced the void formation and growth with increasing RF voltage,
it predicted higher RF voltage of the void formation than it was observed in the experiment.
The ``single-particle'' void was shown to be larger and appear for lower powers than the void with the microparticle suspension effect considered.

Before \cite{Killer-AIPPhysPlas2013}, the researchers mostly aimed at comparing the dimensions and onset of void formation 
in the simulations and the experiment and did not attempt to treat the problem in its entire complexity 
which includes the background plasma along with the microparticle suspension. 
Even the development of particle-in-cell (PIC) codes \cite{Goedheer-ContribPlasPhys2009, Goedheer-JPhysD2009} 
with a more thorough treatment of the electron kinetics did not lead to an immediate breakthrough.
The first attempt \cite{Killer-AIPPhysPlas2013} to compare the outcome of the PIC simulations 
with the microparticle arrangement and RF-period-resolved emission patterns measured simultaneously in the experiment 
was unsuccessful in the sense of void problem since in the simulations,  
``the sharp dust density gradients at the void edges resulted in unwanted boundary effects''.
The PIC model of \cite{Pustylnik-PhysRevE2017} was able to qualitatively reproduce the spatiotemporal emission pattern, 
but was unable to reproduce the formation of void. 
Imposing a fixed configuration of microparticles with a void in the simulation led to tremendous forces on the void boundaries 
and to the drastic distortion of the spatiotemporal emission pattern.

Spectroscopic properties of the stable void were studied in 
\cite{Killer-AIPPhysPlas2013, Tawidian-EPS2013,  Pustylnik-PhysRevE2017, Stefanovic-PlasmaSrcSciTech2017}.
According to \cite{Killer-AIPPhysPlas2013}, the period-resolved emission spectroscopy showed strong increase
of the discharge emission in the center of the discharge, where the void was located.
The void in that experiment was so large that it was not clear if the emission was coupled with the void.
Similar experiments \cite{Pustylnik-PhysRevE2017} with varying the void position showed no void-related feature
in the spatio-temporal emission patterns.

In the discharge image published in \cite{Tawidian-EPS2013}, a bright emission from the void is clearly visible.
Similar feature was also present in the distribution of the resonance atoms measured with laser-induced fluorescence.
The plasma emission and the metastable density distribution were studied in \cite{Stefanovic-PlasmaSrcSciTech2017}
for the argon-acetylene mixture.
The void appeared and grew in size with the growth of the nanoparticles.
Although the plasma emission and the metastable density decreased during the void growth, no local effects could be seen.

In some conditions, the void becomes unstable for high discharge powers and rotates around the discharge axis.
For such experiments, bright emission inside the void (so called ``plasmoid'') was reported in 
\cite{Samsonov-PhysRevE1999, Schulze-PlasmaSrcSciTehnol2006, Lagrange-JApplPhys2015}.
The bright emission from the void was also observed during the void-collapse phase of the heartbeat instability
\cite{Mikikian-NewJPhys2007, Pustylnik-AIPPhysPlas2012}.

The literature review given above shows that there are still lots of unsolved issues regarding the voids in complex plasmas.
State-of-the-art theories explain neither the mechanisms controlling the plasma emission inside the void, nor the void stability. 
The inability to explain the void formation under certain condition implies that mechanisms other than simple balance 
between the electrostatic and ion drag forces may be responsible for the void formation.

In the present work, we tried to make the next step in the understanding of the void problem.
We performed a systematic investigation of capacitively-coupled RF complex plasmas with the void. 
The investigation included measurements of the microparticle configurations, 
time-averaged emission profiles and RF-period-resolved spatiotemporal emission profiles. 
We interpret these data using a simplified and flexible 1D fluid model of a microparticle-containing RF discharge with artificially added radial diffusion. 
Using the experimental data and the model, we were able to demonstrate the limits of applicability of the old void formation mechanism 
(due to a strong ion flow from the void to the suspension) and suggest a novel mechanism explaining the void formation outside those limits.

\section{Experiment}

The experiments were conducted in the ground-based  PK-3 Plus chamber \cite{Thomas-NewJPhys2008}. 
The plasma was produced by means of a capacitively-coupled rf discharge. 
Two electrodes were driven in a push-pull mode by a sinusoidal signal with the frequency of 13.56~MHz. 
Argon fed into the chamber with 3~sccm gas flow was used as a working gas. 
We used melamine formaldehyde spheres with the diameter of $1.95 \pm 0.05 \, \mu$m as microparticles.
The microparticles were injected into the discharge by a electromagnet-driven dispenser through a sieve.
We used the number of the injections as a measure of the microparticle amount in the plasma.

\begin{figure}[htb]
\centering
\includegraphics[width=8cm]{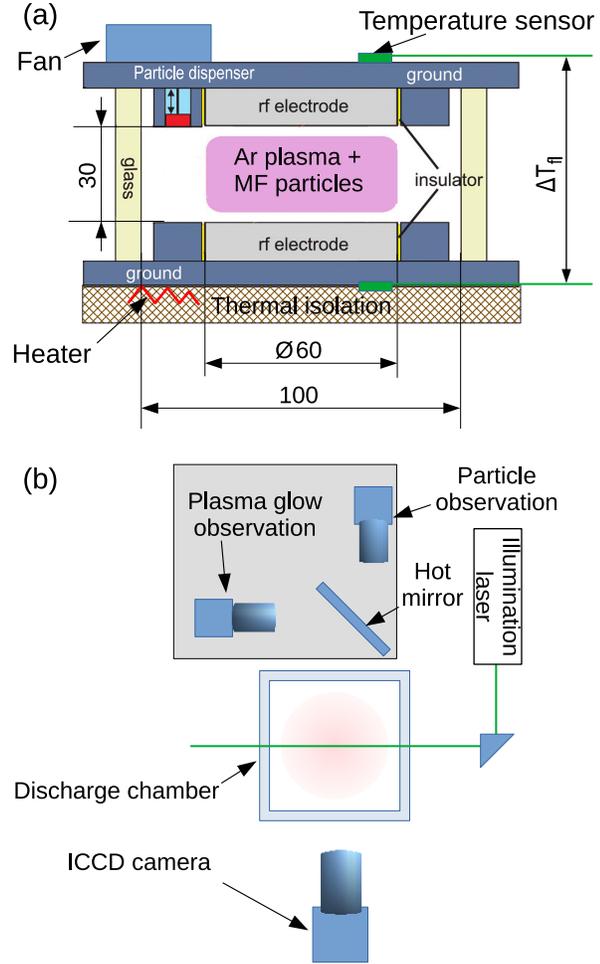}
\caption{Scheme of the experimental setup, (a) --- side view, (b) --- top view. 
The heart of the experiment is the PK-3 Plus chamber \cite{Thomas-NewJPhys2008}.
The bottom flange of the chamber can be heated to control a temperature gradient between the electrodes.
The two cameras on the one side observe the microparticle motion and period-averaged plasma emission.
The ICCD camera on the other side performs RF-period-resolved emission registration.
}
\label{scheme}
\end{figure}

Under ground laboratory conditions, the microparticles concentrate themselves in the vicinity of the lower electrode. 
To obtain large volumetric microparticle suspensions, we compensated the gravitational force by means of themophoresis \cite{Rothermel-PhysRevLett2002}.
The temperature gradient between the electrodes was controlled by heating the bottom electrode with an electric heater 
and cooling the top electrode with fans.
We measured temperatures on the outer surfaces of the two chamber flanges with Tinytag Explorer data logger.
The temperature difference was 14--15~K during the experiments.

The microparticles were illuminated by a laser sheet with the wavelength of 532~nm.
Two Ximea MQ042RG-CM video cameras with interference bandpass filters 
captured the microparticle motion and the plasma emission at the same discharge area.
The filters had the central wavelength of 532 and 750~nm, respectively, and the width of the transmission band of filters was about 10~nm.
Hence, the plasma glow observation camera captured the plasma radiation in 750.4 and 751.5~nm spectral lines.
These lines correspond to the transitions 2p$_{1}$\,$\rightarrow$\,1s$_{2}$ and 2p$_{5}$\,$\rightarrow$\,1s$_{4}$ in the Paschen notation.
The spatial resolution of the cameras was about 37~$\mu$m/pixel.

These diagnostics were supplemented with a RF-period-resolved optical emission spectroscopy, described in details in \cite{Pustylnik-PhysRevE2017}.
RF-period-resolved evolution of the plasma emission was observed by an ICCD camera equipped with a 750~nm central wavelength and 10~nm bandwidth filter.
The 750.4 and 751.5~nm argon lines transmitted by it have the lifetimes of 22.5 ns and 24.9 ns, respectively \cite{Wiese-PhysRevA1989}. 
The camera was synchronized with the rf signal through the frequency divider, 
which sent a synchronization pulse after every 512 RF cycles. 
The gate width of the ICCD camera was set to 10 ns. 
The signal was accumulated over the total exposure time of 1 s. 
The ICCD gate was moved over almost three rf periods with a step of 2 ns.
The spatial resolution of the ICCD camera was about 92~$\mu$m/pixel.

All three cameras were focused on the central cross-section of the discharge chamber.

\begin{figure}[htb]
\centering
\includegraphics[width=7cm]{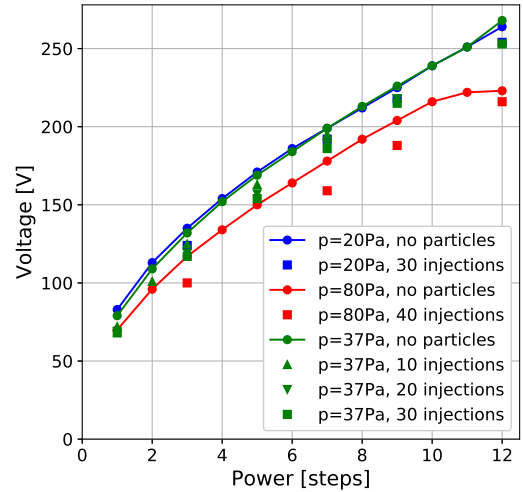}
\caption{Power dependence of the peak-to-peak voltage. }
\label{voltage}
\end{figure}

We injected the microparticles into a low-power (200--300~mW) discharge.
For the suspension stabilization, we waited several minutes, increased the discharge power up to the maximal 
value used in the experiment ($\sim 1.2\,$W)
and decreased it to the minimal value of 100~mW ($\sim 250$~mW for the pressure of 80~Pa).
After that, we observed the microparticle suspension and the plasma emission while
increasing the discharge power in steps.
The power step was slightly pressure dependent.
For the pressure of 20 and 37~Pa it was about 100~mW.
For the pressure of 80~Pa and low discharge power, the step was about 80~mW,
but for the high discharge power, the power measurements were inaccurate due to too high discharge current.
The peak-to-peak voltages between the electrodes for the presented experimental conditions are shown in figure \ref{voltage}.
The values for the pristine plasma are reproducible with an accuracy of 5~V.
The injection of the microparticles caused a drop of the discharge current,
whereas the decrease of the voltage was insignificant.
After every change of the power, we waited 2--3 seconds for the microparticle redistribution.
The plasma emission images were obtained by averaging 20--60 frames after the suspension stabilization.
After reaching the maximal power, we continued the measurements by decreasing the power in the same steps.
The results for increasing and decreasing power were similar and no hysteresis was found, 
therefore we present only the results with the increasing power.

\section{Experimental results}

\begin{figure*}[htb]
\centering
\includegraphics[width=0.8\linewidth]{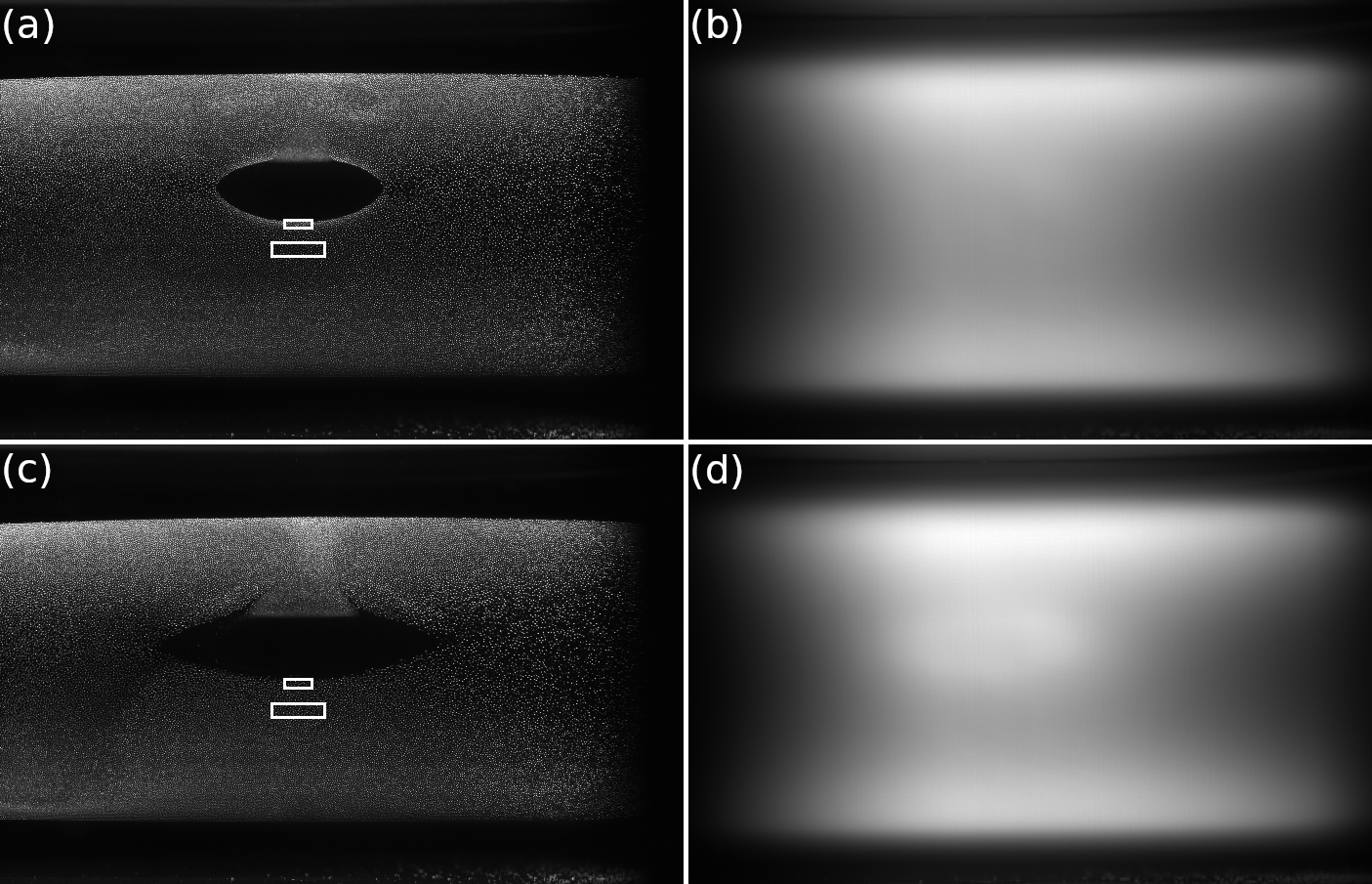}
\caption{Images of the microparticle suspensions (a, c) and the plasma emission (b, d).
The pressure was 37~Pa, the discharge power was (a)--(b) 5 steps, (c)--(d) 7 steps (see figure~\ref{voltage}).
The white rectangles depict the areas used for figure \ref{chart-void}(c). 
}
\label{real}
\end{figure*}

For low power, no void was present.
The onset of the void formation was pressure dependent: step 1 for the pressure of 20~Pa, step 2 for 37~Pa and step 3 for 80~Pa.
Within several power steps after the void formation, no void-related feature could be seen in the plasma emission.
In the following, we will call this regime ``dim void'' (see figures \ref{real}(a--b)).
After the power reached a certain threshold, bright emission associated with the void area could be observed.
In the following, we will call this regime ``bright void'' (see figures \ref{real}(c--d)).

\begin{figure*}[htb]
\centering
\includegraphics[width=0.9\linewidth]{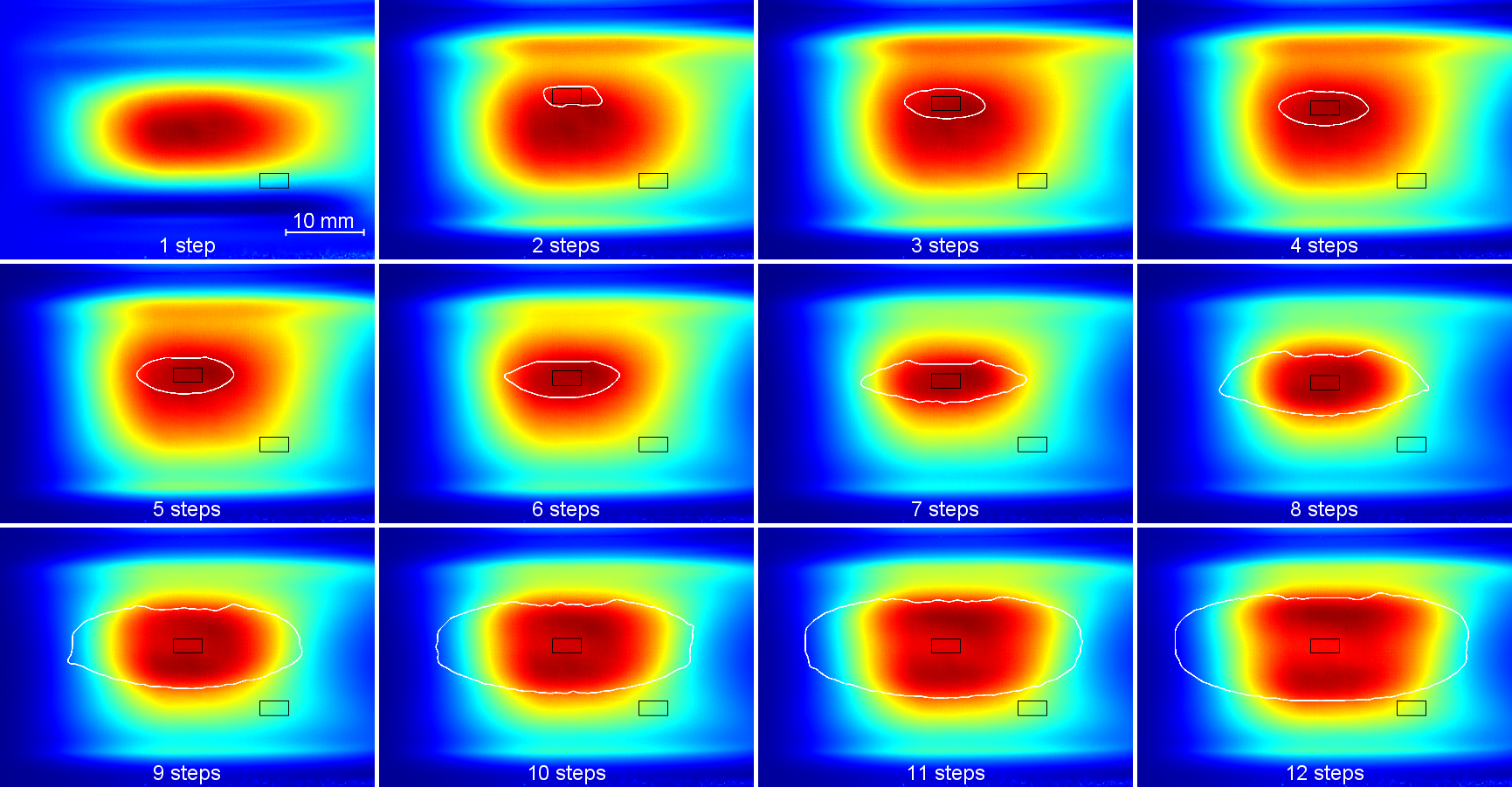}
\caption{The colour-coded effect of the microparticles (30 injections) on the plasma emission with the filter for 750~nm. 
Pressure --- 37~Pa, power --- 1--12 steps (see figure~\ref{voltage}).
Every image has its own colour scale.
The white lines depict the void boundaries. 
The black rectangles show the areas used for figure \ref{chart-intensity}.
}
\label{dusty3-full}
\end{figure*}

\begin{figure*}[htb]
\centering
\includegraphics[width=0.675\linewidth]{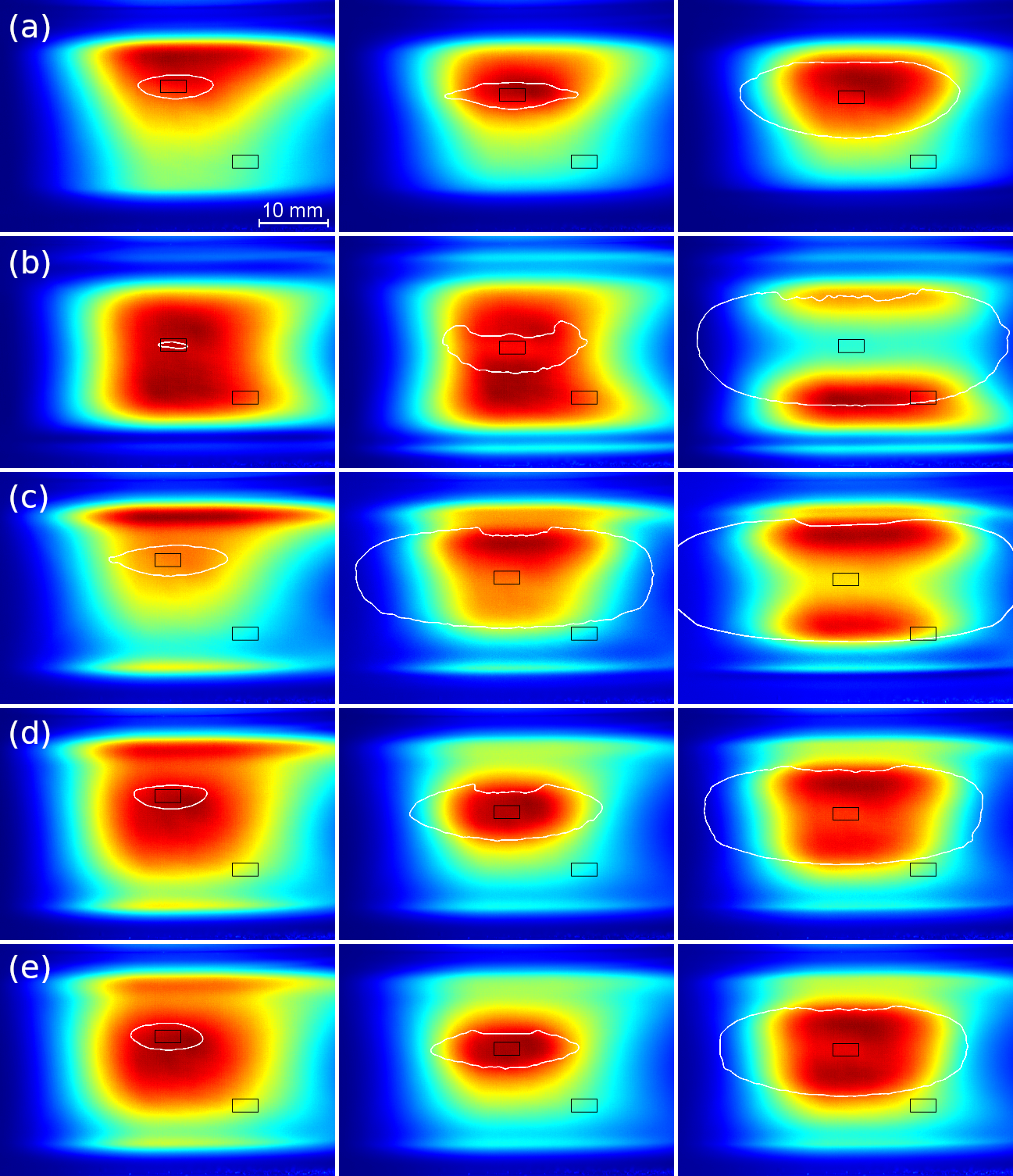}
\caption{The same as figure \ref{dusty3-full} for the power of (from left to right) 3, 7 and 11 steps (see figure~\ref{voltage}) and
(a) pressure 20~Pa, 30 injections;
(b) 80~Pa, 40 injections;
(c) 37~Pa, 10 injections;
(d) 37~Pa, 20 injections;
(e) 37~Pa, 30 injections.}
\label{camB-compare}
\end{figure*}

To make the effect of the microparticles more pronounced, we subtracted the emission of the pristine plasma from the images with the microparticles.
The results are presented in figures \ref{dusty3-full}--\ref{camB-compare}.

The boundaries of the voids shown in figures \ref{dusty3-full}--\ref{camB-compare} were found 
as the boundaries of the areas in the microparticle suspension images with 
the intensity less than a certain threshold.
Using these boundaries, the void dimensions (height and width) were determined. 
To mitigate the influence of small-scale roughness of the boundaries on the void dimensions, 
the height and width of the void were determined as, repectively, the height and width of such a rectangle 
that each of its sides cuts off four percent of the total amount of the points belonging to the boundary.
Also, we measured the effect of the microparticles on the plasma emission intensity inside the void area and outside the void.

\begin{figure}[htb]
\centering
\includegraphics[width=8cm]{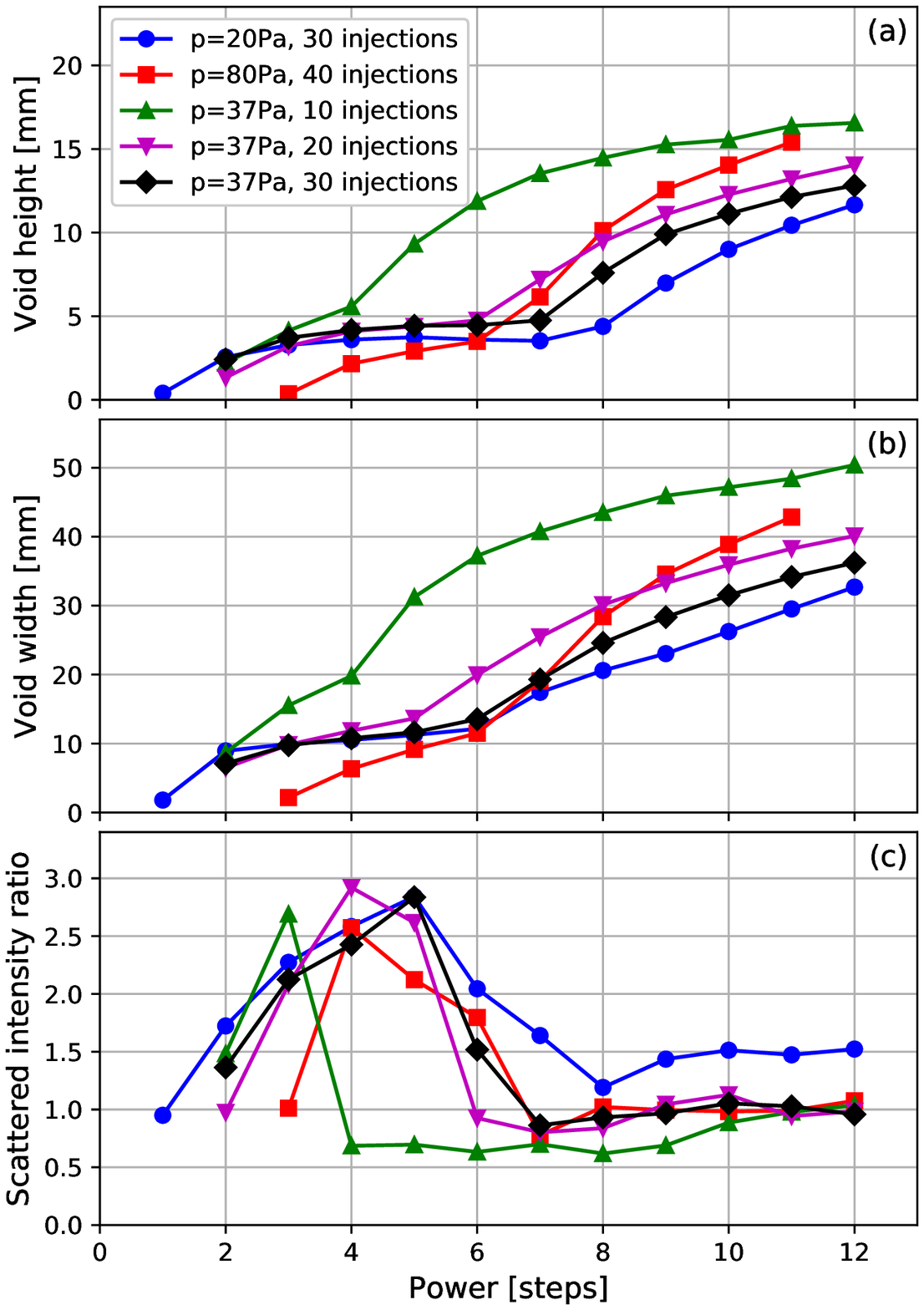}
\caption{ The parameters of the void for different discharge conditions.
(a) --- the void height; 
(b) --- the void width;
(c) --- the ratio of the 532~nm light intensity scattered on the microparticles at the bottom void border and in the bulk suspension.
}
\label{chart-void}
\end{figure}

\begin{figure}[htb]
\centering
\includegraphics[width=8cm]{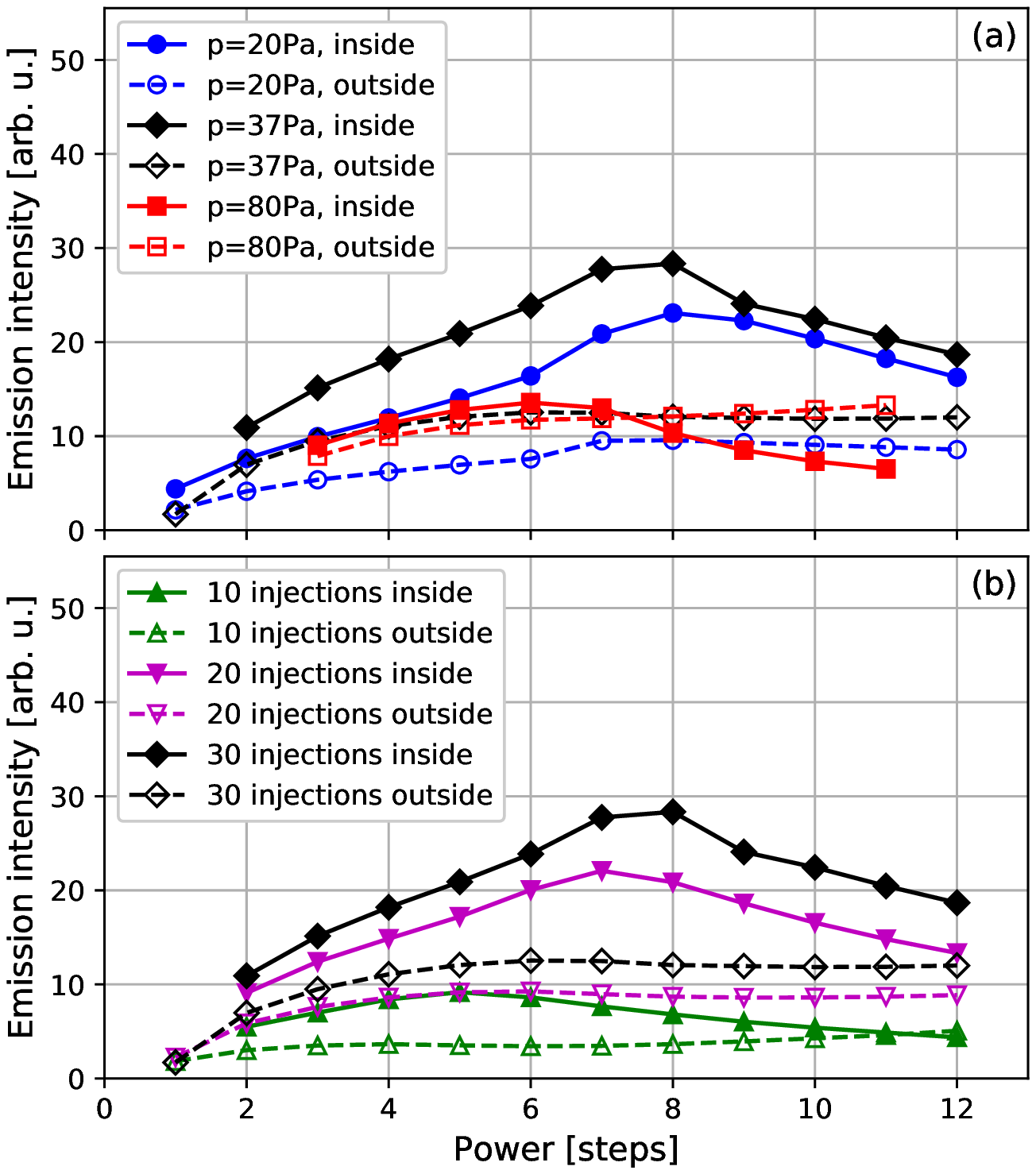}
\caption{ 
The microparticle effect on the plasma emission (in arbitrary units) inside and outside the void area.
(a) --- for different gas pressures; 
(b) --- for the gas pressure of 37~Pa and different amounts of the microparticles.
}
\label{chart-intensity}
\end{figure}

The data are presented in figures \ref{chart-void}--\ref{chart-intensity}.
The areas of the images used to measure the plasma emission are depicted in figures \ref{dusty3-full}--\ref{camB-compare} with the black rectangles.
Due to imperfect gravity compensation, the void was located not in the center of the chamber, and its position was dependent on the discharge power.
That is why we adjusted the position of the measurement area inside the void for every power step.
The position of the area outside the void was fixed.

The experiments showed that, after the void formation, its size stabilized, 
and the void was growing only slightly with the power increase, similarly to \cite{Lipaev-PhysRevLett2007, Schmidt-AIPPhysPlas2011, Sarkar-PlasmaSrcSciTech2015}.
However, after the power reached a certain value, the horizontal void size started increasing faster.
After that, the bright plasma emission appeared inside the void and the void height started growing faster.
The void, therefore, experienced the transition from dim to bright regime.
For the large voids, the microparticles cause the plasma emission increase concentrated near the top and the bottom void boundaries.
This effect is more pronounced for higher gas pressure.

We also noticed a transformation of the microparticle arrangement on the void boundary during this transition.
At the dim void boundary, the microparticle suspension was denser than in the bulk,
whereas for the bright void boundary, no density peak was observed, 
similar to the measurements reported in \cite{Lipaev-PhysRevLett2007}.
The effect is qualitatively visible in figures \ref{real} (a, c).
To quantify it, we calculated the ratio of the averaged scattered light intensity 
(assuming it to be proportional to the microparticle number density) in a small rectangle at the bottom void boundary 
and in a rectangle in the microparticle suspension bulk (see  figure \ref{chart-void} (c)).
Examples of such rectangles are shown in figures \ref{real}.
For every image, we measured these intensities 3 times slightly varying the position of the areas and averaged the results.

\begin{figure}[htb]
\centering
\includegraphics[width=6cm]{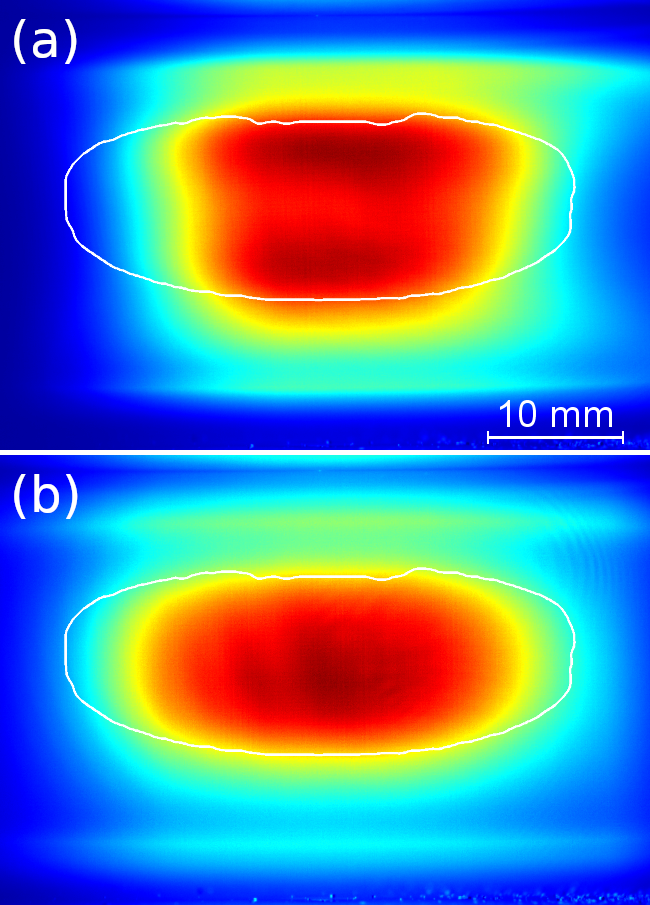}
\caption{The colour-coded effect of the microparticles on the plasma emission with the filter for (a) 750~nm, (b) 810~nm. 
Pressure --- 37~Pa, power --- 12 steps (see figure~\ref{voltage}), 30 injections of the microparticles.
}
\label{compare}
\end{figure}

We also observed the plasma emission through a narrow-band filter with 810~nm central wavelength,
which transmitted 810.4 and 811.5~nm spectral lines.
The effect of the microparticles on the plasma emission in these lines is similar,
but no sheaths on the void boundaries were observed even for the largest power 
(see figure \ref{compare}).

\begin{figure*}[htb]
\centering
\includegraphics[width=1\linewidth]{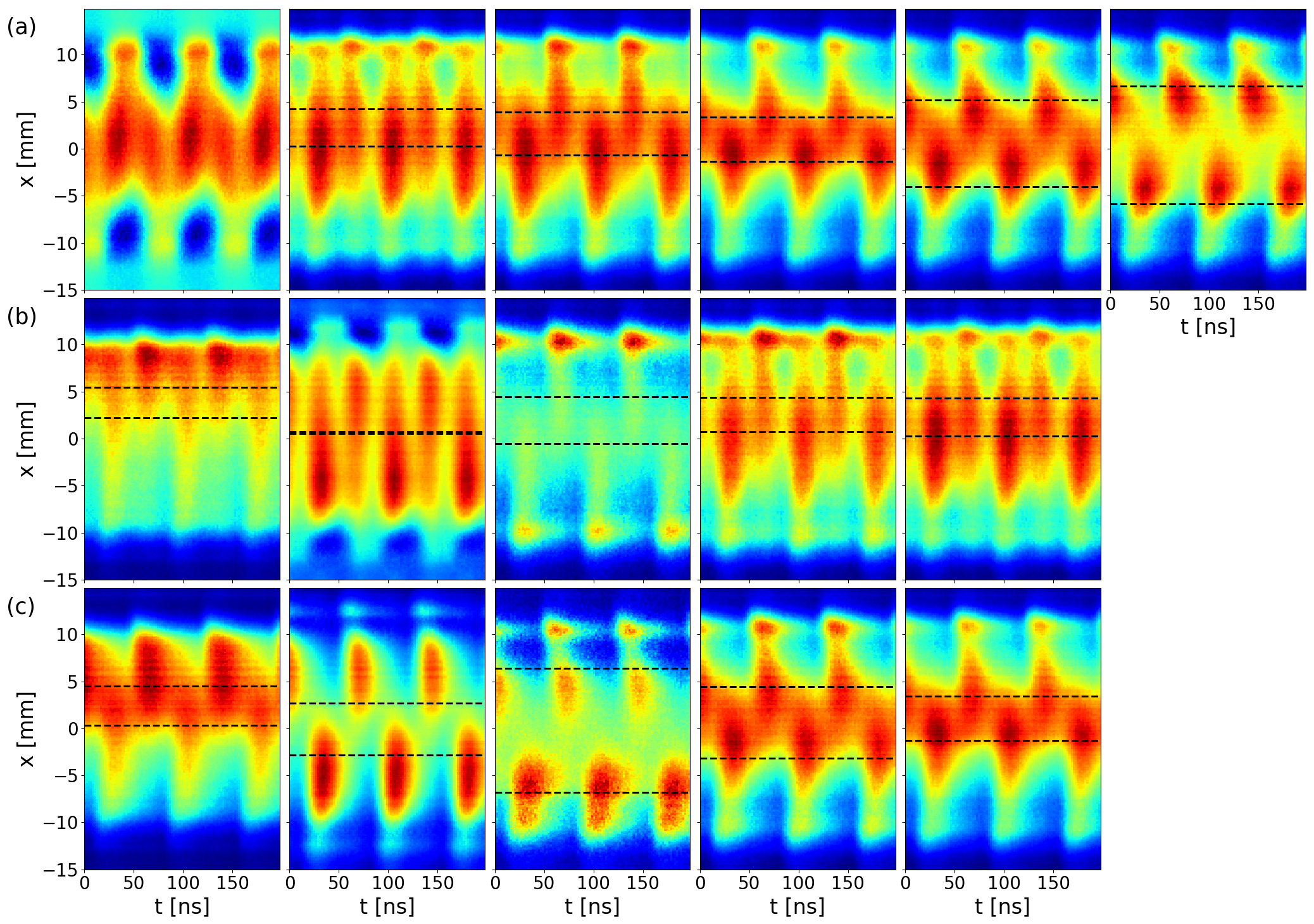}
\caption{Spatiotemporal patterns of the microparticle effect on the plasma emission in the center of the discharge image
for (a) pressure 37 Pa, 30 injections, power values (from left to right) of 1, 3, 5, 7, 9 and 12 steps; 
(b) power of 3 steps and (from left to right) pressure 20 Pa and 30 injections, 80 Pa and 40 injections, 
37 Pa and 10 injections, 37 Pa and 20 injections, 37 Pa and 30 injections and 
(c) power 7 steps and same pressures and microparticle amounts as for plate (b).
$x$ is the coordinate along the discharge axis with $x=0$ in the discharge center.
Every image was colour-coded individually.
The dashed lines depict the void boundaries.
In plate (a), the microparticles caused decrease of the plasma emission near the expanding sheath for the lowest power.
Correspondence of power steps and rf peak-to-peak voltage is shown in figure~\ref{voltage}.
}
\label{period}
\end{figure*}

The results of the RF-period-resolved optical emission spectroscopy are presented in figure \ref{period}.
Every pixel column for this spatiotemporal pattern was obtained by averaging the intensities
along the horizontal dimension a central part of one frame.
The width of the averaged area was about 6~mm.
The images in figure \ref{period} are the difference between the patterns with and without the microparticles for the respective conditions.

The results of the RF-period-resolved spectroscopy are in accord with the RF-period-averaged results.
In the case of low-power discharge, the effect of the microparticles is significant 
for entire discharge height every half period.
It leads to a more uniform spatial distribution of the emission intensity, similar to \cite{Pustylnik-PhysRevE2017}.
For the lowest power, the microparticles cause the emission increase near collapsing sheath,
but smaller effect or even decrease of the emission near expanding sheath, as reported in \cite{Killer-AIPPhysPlas2013}.

With the increase of power, the effect appears only inside the void and between the void and the expanding sheath.
With further increase of the power, the emission increase concentrates near the void boundary 
that corresponds to the electrode with the expanding sheath.

\begin{figure}[htb]
\centering
\includegraphics[width=7cm]{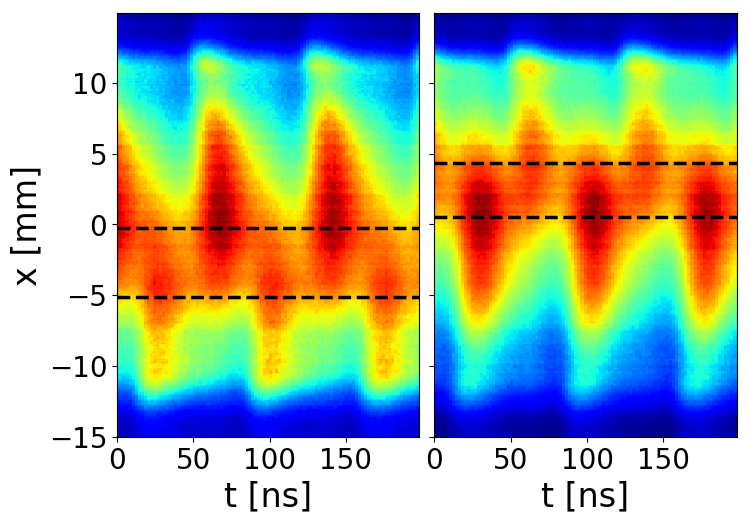}
\caption{Spatiotemporal patterns of the microparticle effect on the plasma emission for different void position
for the pressure of 37~Pa and the power of 7 steps (see figure~\ref{voltage}).
}
\label{position}
\end{figure}

We varied the vertical position of the bright void by changing the temperature difference between the electrodes
while keeping the discharge power, gas pressure and microparticle amount constant.
The results are shown in figure \ref{position}. 
We see that the emission pattern moves together with the void. 
Therefore, the emission profile observed in the bright void regime is really void-related.

\section{Discussion}
\label{sec: Disc}
\subsection{Time-averaged 1D fluid model}
\label{sec: Mod}
To provide a qualitative analysis of the experimental results, we consider a simple one-dimensional fluid  model that describes time-averaged profiles of the electron, ion and dust number densities near the discharge axis. The model is applied to analyse possible configurations of microparticle suspensions and corresponding distributions of the electric field in the discharge. This information is then used to discuss the emission patterns observed experimentally. \\ \indent
Let $x$ denote the discharge axis. We assume that $x \in [0,L]$, where $x=0$ is the point in the middle of the discharge gap and $x=L$ is the point at the electrode surface ($L=15\,$mm).
The distribution of the ion density, $n_{i}$, is governed by the continuity equation:
\begin{equation}
\label{EqNi}
\frac{\partial}{\partial x}(n_{i}u_{i})=G_{ion}-G_{d},
\end{equation}
where $u_{i}$ is the ion velocity, $G_{ion}$ is the ionization source term, $G_{d}$ is the sink term due to absorption of ions by the microparticles. The ion flux in equation~(\ref{EqNi}) is approximated as 
\begin{equation}
\label{EqJi}
n_{i}u_{i}=n_{i}\bar{u}_{i}-D_{i}{\partial n_{i}}/{\partial x},
\end{equation}
where $D_{i}=0.012\,$m$^2$s$^{-1}$ is the ion diffusion coefficient and $\bar{u}_{i}$ is the ion drift velocity defined as a function of the local electric field, $E$. The ion drift velocity is obtained by solving the equation $eE=\omega_{ia}(\bar{u}_{i})m_{i}\bar{u}_{i}$, where $m_{i}$ is the ion mass and
$\omega_{ia}$ is the effective ion-atom collision frequency defined as \cite{brinkmann2011plasma}: $\omega_{ia}=\sigma_{0}n_{g}\sqrt{u_{0}^{2}+\bar{u}_{i}^{2}}$, where $n_{g}$ is the number density of gas atoms, $\sigma_{0}=10^{-18}\,$m$^{2}$, $u_{0}=550\,$m\,s$^{-1}$ (here $n_{g}=p_{g}/k_{\rm B}T_{g}$, where $T_{g}=300\,$K and $p_{g}$ is the gas pressure). This approximation agrees well with the Monte-Carlo simulation data for moderate drift velocities \cite{semenov2017moment}. For simplicity, we neglect the effect of microparticles on the ion drift flow. The estimates show that this assumption is acceptable for the conditions of the present study.
\\ \indent
The electron density, $n_{e}$, is assumed to obey the Boltzmann relation, which can be written in the following form:
\begin{equation}
\label{EqNe}
\mu_{e}En_{e}=-D_{e}{\partial n_{e}}/{\partial x},
\end{equation}
where $\mu_{e}$, $D_{e}$ are the electron mobility and diffusion coefficient, respectively.\\ \indent
The microparticle number density, $n_{d}$, is governed by the force balance:
\begin{equation}
\label{EqNd}
-\frac{\partial p_{d}}{\partial x}+n_{d} \left( q_{d} E + F_{id} + F_{th}\right)=0,
\end{equation}
where $q_{d}$ is the microparticle charge, $p_{d}$ is the excess pressure of microparticles, $F_{th}$ is the thermophoretic force associated with the internal heat sources of the plasma and $F_{id}$ is the drag force exerted by ions on a single microparticle (the ion drag force). 
The gravity force is assumed to be compensated in the vertical direction (see ~\ref{sec: DustPlasm}).
The excess pressure in equation~(\ref{EqNd}) can be calculated using the existing thermodynamic models \cite{khrapak2014ion,khrapak2014simple,khrapak2015practical}.
In the present work the model of~\cite{khrapak2015practical} was used. The expression for the excess pressure reads:
\begin{equation}
\label{Pd}
p_{d}=n_{d}k_{\rm B}T_{g}\left \{ 1+ \frac{\Gamma \kappa^{4}}{6[\kappa\cosh(\kappa)-\sinh(\kappa)]^{3}}\right \},
\end{equation}
where $\kappa$ and $\Gamma$ are the screening and coupling parameters, respectively. The screening and coupling parameters are given by $\kappa=a/\lambda_{s}$ and $\Gamma=q_{d}^{2}/\Delta k_{\rm B} T_{g}$, where 
$\Delta=(3/4 \pi n_{d})^{1/3}$ is the Wigner-Seitz
radius and $\lambda_{s}$ is the screening length of the interparticle interaction potential (we assume that particles interact via Yukawa potential). 
The microparticle density profile given by equation~(\ref{EqNd}) is assumed to satisfy the integral constraint
\begin{equation}
\label{Ind}
\int_{0}^{L} n_{d} dx=\bar{n}_{d}L,
\end{equation}
where $\bar{n}_{d}$ is the averaged microparticle density.\\ \indent
Equations~(\ref{EqNi}),~(\ref{EqNe}) and (\ref{EqNd}) are combined with the Poisson equation for the electric field:
\begin{equation}
\label{EqPsn}
\frac{\partial E}{\partial x}=\frac{e}{{\varepsilon_{0}}}\left( n_{i}-n_{e}-Z_{d}n_{d}\right),
\end{equation}
where $Z_{d}=|q_{d}|/e$, with $q_{d}$ being the microparticle charge.
The microparticle charge is given by $q_{d}=4\pi\varepsilon_{0}a_{d}\varphi_{d}$, where $a_{d}$, $\varphi_{d}$ are the particle radius and potential, respectively.
\\ \indent
In order to close the model, it is necessary to define  $G_{ion}$, $G_{d}$, $F_{th}$ and $F_{id}$ in equations (\ref{EqNi}) and (\ref{EqNd}). The interaction of microparticles with the plasma components is relatively well studied and reliable expressions for $G_{d}$, $F_{th}$, $F_{id}$ are available. The details on how these terms are calculated in the present work are given in~\ref{sec: DustPlasm}.\\ \indent
A self-consistent calculation of $G_{ion}$ in equation~(\ref{EqNi}) would require consideration of the electron energy balance and electron dynamics in the RF-discharge. This goes beyond the goal of the present analysis. Thus, we use $G_{ion}$ as an input parameter of the model to study possible equilibrium configurations of the dust component.  Particularly, we use the following model profile of the ionization source term:
\begin{equation}
\label{Gion}
G_{ion}=G_{0}\left[\theta+(1-\theta)e^{-(x-x_{0})^{2}/r_{0}^{2}}\right],
\end{equation}
where $r_{0}=3\,$mm, $x_{0}=10\,$mm, $G_{0}$ and $\theta$ are adjustable parameters ($G_{0} \sim 10^{20}\,$m$^{-3}$s$^{-1}$, $\theta \lesssim 0.5$). 
The choice of the ionization source profile~(\ref{Gion}) was motivated by the results of particle-in-cell simulations performed for similar conditions \cite{Pustylnik-PhysRevE2017}.\\ \indent
The boundary conditions for equations~(\ref{EqNi})-(\ref{EqPsn}) are as follows. We set $\partial n_{s}/\partial x=0$ ($s=e,i,d$) and $E=0$ at $x=0$ and we use $\partial n_{i}/\partial x=0$, $n_{d}=0$, $n_{e}=\bar{n}_{e}$ at $x=L$. Equations~(\ref{EqNi})-(\ref{EqPsn}) with the boundary conditions were discretized using the finite-difference method and solved iteratively (the problem was reduced to a boundary-value problem for the governing equations).\\ \indent
The computations were performed for the following conditions: $a_{d}=1\,\mu$m, $\varphi_{d}=-4\,$V, $p_{g}=37\,$Pa. The averaged electron density at the electrode surface was set to $\bar{n}_{e}=10^{13}\,$m$^{-3}$ (the modelling results are quite insensitive to variations of $\bar{n}_{e}$). The electron temperature, $T_{e}$, was assumed to be constant throughout the discharge and is set to 4\,eV. The normalized microparticle potential, $e|\varphi_{d}|/k_{\rm B}T_{e}$, is equal to one in this case (this is a typical value for the conditions of our study). The electron mobility and diffusion coefficient were calculated using Bolsig+ code \cite{hagelaar2005solving} at a given electron temperature. The cross-sections from Biagi database were used \cite{biagi}.
The effect of the microparticles on the electron distribution function was neglected. The above conditions were chosen to match the experimental conditions and previous results of particle-in-cell simulations\cite{Pustylnik-PhysRevE2017}. The other parameters of the model are specified separately for the results presented  in subsection~\ref{subsec: SimRes}.\\ 

\subsection{Plasma emission}

In order to qualitatively assess the phase-resolved emission pattern in the bulk plasma, we consider the simplified electron energy equation:
\begin{equation}
\label{EqUe}
\frac{\partial \varepsilon_{e}}{\partial t}=\mu_{e}(\varepsilon_{e})\bar{E}(x,t)^{2}-S_{e}(\varepsilon_{e}),
\end{equation}
where $t$ is time, $\varepsilon_{e}(x,t)$ is the mean electron energy and $S_{e}$ is the energy losses due to elastic and inelastic processes (computed using Bolsig+ solver). The electric field in equation~(\ref{EqUe}) is approximated as
\begin{equation}
\label{EqUeE}
\bar{E}(x,t)=E(x)+E_{0}\sin(\omega t),
\end{equation}
where $E(x)$ is the time-averaged field given by equation~(\ref{EqPsn}) and $E_{0}$ is the amplitude of the oscillating rf field (here, $\omega/2 \pi =13.56\,$MHz). The amplitude of the oscillating field was set to $E_{0}=1\,$V\,mm$^{-1}$ (characteristic value that is used in all computations). Certainly, equation~(\ref{EqUe}) cannot be used close to the electrodes. Nevertheless, for the bulk plasma, this simplified model can be expected to provide a qualitatively correct description of the electron energy variation. Equation~(\ref{EqUe}) was solved numerically using the second-order Runge-Kutta method.\\ \indent
Using the solution of equation~(\ref{EqUe}), we estimate the time-space dependence of the total excitation rate for the 2p$_1$ and 2p$_5$ argon levels, 
which are the upper levels of the 750.4 and 751.5 nm transitions used in the experiment for the RF-period-resolved measurements.
The excitation rate is given by $R_{ex}=k_{ex}(\varepsilon_{e})n_{g}n_{e}$, where the rate coefficient $k_{ex}$ is calculated using the Bolsig+ solver. The emission pattern was calculated by the convolution of the total excitation rate and the decay function $\exp(-t/\tau)$, with $\tau=23\,$ns being the characteristic lifetime of the transition.

\subsection{Modelling results}
\label{subsec: SimRes}
Before discussing the modelling results, it is helpful to consider the balance between the electric force ($F_{e}=q_{d}E$) and the ion drag force acting on a single microparticle. 
The balance between these two forces is supposed to play a major role in determining the microparticle suspension configuration.
Since the ion velocity in the discharge is usually close to the local drift velocity for the considered conditions \cite{semenov2017moment}, 
both forces can be considered as functions of the uniform electric field causing the respective mobility-limited ion flow.

\begin{figure}
\centering
\includegraphics[width=0.4\textwidth]{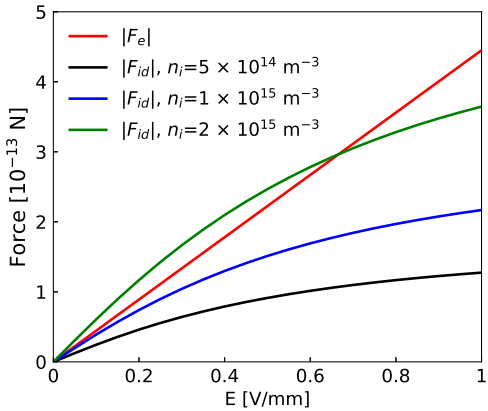}
\caption{\label{fig: drag} The electric force ($F_{e}$) and the ion drag force ($F_{id}$) acting on a single microparticle for the case of the ion drift flow. The ion drag force is shown for typical values of the ion density.}
\end{figure}

In figure~\ref{fig: drag}, we show the electric force and the ion drag force computed as functions of the electric field, where the ion drag force is presented for typical values of the background ion  density (under the assumption that $n_{i}=n_{e}$). As seen in figure~\ref{fig: drag}, there are regimes where the equilibrium point ($|F_{e}|=|F_{id}|$) exists, and where $|F_{e}|>|F_{id}|$ for the entire considered range of electric fields. The existence of
the equilibrium point is usually associated with the formation of a central void in the microparticle suspension. In the opposite regime (where $|F_{e}|>|F_{id}|$), the microparticles are expected to be uniformly distributed in the discharge \cite{Pustylnik-PhysRevE2017}.

\begin{figure*}
\centering
\includegraphics[width=1\textwidth]{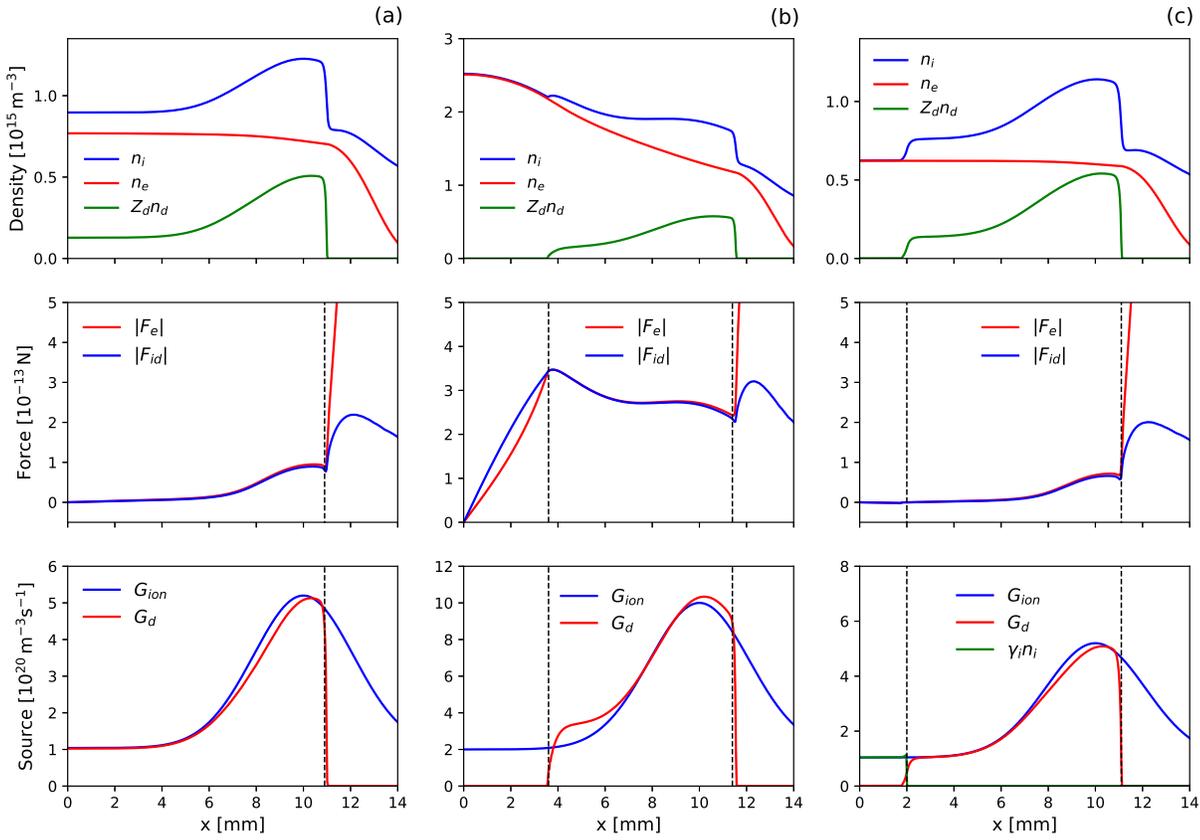}
\caption{\label{fig: cloud} The computation results for $G_{0}=5.2 \times 10^{20}\,$m$^{-3}$s$^{-1}$ (a), $G_{0}=10 \times 10^{20}\,$m$^{-3}$s$^{-1}$ (b) and $G_{0}=5.2 \times 10^{20}\,$m$^{-3}$s$^{-1}$, $\gamma_{i}=1.67\times10^{5}$s$^{-1}$ (c) : density profiles, distribution of the electric and ion drag forces, distribution of the ionization and sink terms. The dashed lines show the boundaries of the microparticle suspension.}
\end{figure*}

Consider, first, the case when the condition $|F_{id}|>|F_{e}|$ is not satisfied in the bulk plasma. If the averaged dust density is set to $\bar{n}_{d}=7\times10^{10}\,$m$^{-3}$, this regime is reproduced by the model for $G_{0} < 6.5\times10^{20}$\,m$^{-3}\,$s$^{-1}$ at $\theta=0.2$ (these values are taken as an example). In figure~\ref{fig: cloud}(a), we present the results of computations for $G_{0}=5.2 \times 10^{20}\,$m$^{-3}$s$^{-1}$. As shown in figure~\ref{fig: cloud}(a), the microparticles are distributed almost uniformly. The suspension is confined by the electric field in the sheath region. In the central region, the production of ions due to ionization is balanced by the absorption of ions on the microparticles ($G_{ion} \approx G_{d}$). As a result, the ion velocity and averaged electric field are almost zero here. Qualitatively, this picture corresponds to the result of previous simulations obtained using one-dimensional models for similar conditions \cite{Pustylnik-PhysRevE2017,sukhinin2013dust}.

Consider now the case when the condition  $|F_{id}|>|F_{e}|$ is satisfied in the bulk plasma. In figure~\ref{fig: cloud}(b), we present the results obtained for $G_{0}=10\times10^{20}\,$m$^{-3}$s$^{-1}$. As it is seen in figure~\ref{fig: cloud}(b), the central void is reproduced by the model.
The microparticle configuration is mainly determined by the balance between $F_{e}$ and $F_{id}$. Here, $|F_{id}|>|F_{e}|$ in the void region and $|F_{e}|>|F_{id}|$ in the sheath region. These two regions provide the confinement of the microparticle suspension. The equilibrium condition, $|F_{d}|\approx |F_{e}|$, holds within the microparticle suspension.
\\ \indent
The estimated emission patterns for the results shown in figure~\ref{fig: cloud}(a--b) are presented in figures~\ref{fig: ems}(a),\,(d), respectively. In addition, in figure~\ref{fig: ems}(c), we show the emission pattern for the intermediate case taken at  $G_{0}=7.5 \times 10^{20}\,$m$^{-3}$s$^{-1}$. The electric field distributions corresponding to the results presented in figure~\ref{fig: ems} are shown in figure~\ref{fig: field}.
\\ \indent
The emission pattern in figure~\ref{fig: ems}(a) is almost uniform and correlates with the experimental observations at low discharge powers, when no void is present. The uniformity of the emission pattern is explained by the fact that the time-averaged electric field is almost zero in the bulk plasma.
The results shown in figures~\ref{fig: cloud}(a) and \ref{fig: ems}(a) are in qualitative agreement with the results of particle-in-cell simulations presented in \cite{Pustylnik-PhysRevE2017}.
The emission pattern in figure~\ref{fig: ems}(d) correlates with the experimental observations at elevated discharge powers (the case of a bright void). The maximum of the emission intensity in the bulk region is located at the void boundary. This correlates with the corresponding distribution of the electric field shown in figure~\ref{fig: field}.\\ \indent

\subsection{The effect of radial diffusion}

\begin{figure*}
\centering
\includegraphics[width=1\textwidth]{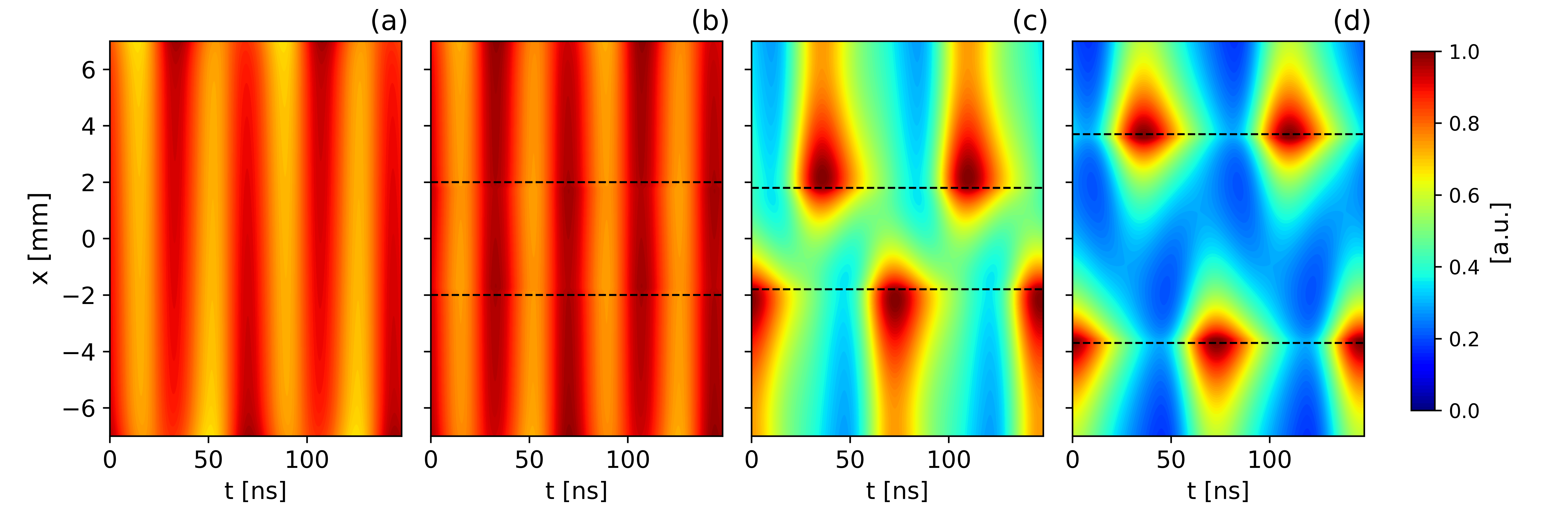}
\caption{\label{fig: ems} The simulated spatio-temporal emission patterns in the bulk plasma region. The results are presented for different values of the parameter $G_{0}$: (a, b) $5.2 \times 10^{20}\,$m$^{-3}$s$^{-1}$, (c) $7.5 \times 10^{20}\,$m$^{-3}$s$^{-1}$, (d) $10 \times 10^{20}\,$m$^{-3}$s$^{-1}$. 
The pattern (b) was obtained taking into account the radial ion diffusion, $\gamma_{i}=1.67\times10^{5}$s$^{-1}$.
The dashed lines show the positions of the void edges.}
\end{figure*}

\begin{figure}
\centering
\includegraphics[width=0.35\textwidth]{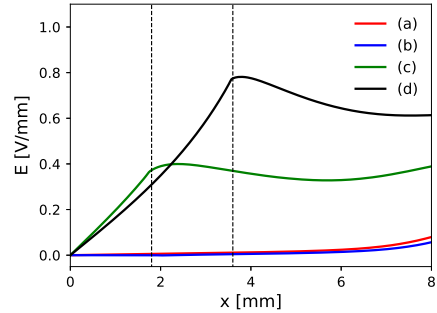}
\caption{\label{fig: field} Distributions of the time-averaged electric field for the emission patterns presented in figures~\ref{fig: ems}(a),\,(b),\,(c) and (d), respectively. The dashed lines show the position of the void edge.}
\end{figure}

The emission pattern in figure~\ref{fig: ems}(c) could probably explain the experimental observations in the case of a dim void. In fact, the electric field is almost uniformly distributed within the microparticle suspension in this case. Thus, the emission intensity at the void boundary is approximately at the same level as in the cloud. Moreover, the experimental picture is obviously affected by the emission from the outer discharge regions, where the microparticles are distributed uniformly (as it was discussed in~\cite{Pustylnik-PhysRevE2017}, this argumentation should be taken with care).\\ \indent
At the same time, the emission pattern observed for the case of a dim void is similar to that shown in figure~\ref{fig: ems}(a). The emission pattern is not noticeably affected by the presence of the void in this regime. This observation is similar to that reported previously in~\cite{Pustylnik-PhysRevE2017}. Thus, it is questionable that the results in figure~\ref{fig: ems}(c) can reliably explain the experimental observations for the case of a dim void.  This, in turn, indicates that there exists another mechanism of the void formation at low discharge powers.  Below, we discuss the results which could provide a hint to such a mechanism.

First, it should be noted that 1D model described in subsection~\ref{sec: Mod} does no take into account the loss of ions due to radial diffusion. At the same time, the rate of this process can be comparable to that of ionization and absorption of ions by the microparticles. In fact, the second order expansion of the ion density near the discharge axis can be written as $n_{i}(r)=n_{i0}\left[1-\delta(r/r_{0})^{2}\right]$, where $n_{i0}$ is the on-axis ion density, $r$ is the radial coordinate, $r_{0}$ is the characteristic length in the radial direction and $\delta$ is the relative density drop. For pure plasma, the loss rate due to radial diffusion can be estimated as $(D_{a}/r)d/dr(rdn_{i}/dr)_{r \rightarrow 0} \approx -4D_{a}\delta n_{i0}/r_{0}^{2}$, where $D_{a} \approx \mu_{i}D_{e}/\mu_{e}$ is the ambipolar diffusion coefficient (here $\mu_{i}=e/(m_{i}n_{g}\sigma_{0}u_{0})$ is the ion mobility). For the conditions of our study we get $D_{a} \approx 4\,$m$^{2}\,$s$^{-1}$. Taking, for example, $r_{0}=5$\,mm (characteristic radius of the dim void), $n_{i0}=10^{15}\,$m$^{-3}$ and $\delta=16 \%$ we get $4D_{a}\delta n_{i0}/r_{0}^{2} \approx 10^{20}\,$m$^{-3}\,$s$^{-1}$. This value is comparable with typical values of $G_{ion}$ and $G_{d}$ in figure~\ref{fig: cloud}(a).\\ \indent
In order to assess the possible effect of radial ion diffusion on the microparticle arrangement, we consider the modified ion continuity equation:
\begin{equation}
\label{EqNimod}
\frac{\partial}{\partial x}(n_{i}u_{i})=G_{ion}-G_{d}-\gamma_{i}n_{i},
\end{equation}
where $\gamma_{i}n_{i}$ simulates the loss of ions due to radial diffusion. The remaining equations of the model are the same as described in section~\ref{sec: Mod}.
\\ \indent
As it was discussed above, $\gamma_{i}n_{i}$ can be of the same order of magnitude as $G_{ion}$ in pure (microparticle-free) plasma. At the same time, it can be expected that the radial losses within the microparticle suspension are negligible (or at least lower than in dust-free plasma). The general trend observed in different simulations is that $G_{ion}\approx G_{d}$ in the microparticle suspension for the considered conditions (see figures~\ref{fig: cloud}(a--b), and the results in  
\cite{Pustylnik-PhysRevE2017,sukhinin2013dust}). Thus, the ion velocity can become relatively low within the suspension (see figure~\ref{fig: cloud}(a), here $F_{id} \sim u_{i}$). Thus, it seems reasonable to assume that $\gamma_{i}=0$ inside the microparticle suspension. 
\\ \indent
These two observations can be summarized as follows. If we assume, that there is a microparticle-free region $x \le x_{v}$, where $\gamma_{i} \ne 0$ and $G_{ion}=\gamma_{i}n_{i}$, then $n_{d}$ can become zero in this region, i.e., the void is formed. It is remarkable that our model can indeed reproduce such a configuration. In figure~\ref{fig: cloud}(c), we show the results obtained for $G_{0}=5.2 \times 10^{20}\,$m$^{-3}$\,s$^{-1}$, $\gamma_{i}=1.67\times10^{5}$\,s$^{-1}$ and $x_{v}=2\,$mm ($\gamma_{i}=0$ for $x > x_{v}$). The solution shown in figure~\ref{fig: cloud}(c) represents a stable configuration with the microparticle-free region in the center. The electric force is positive in the void and dominates over the negative ion drag force. This explains the stability of the void in this regime.\\ \indent
In addition, it should be emphasized that the electric force (and consequently the electric field) at the void boundary is lower than that shown in figure~\ref{fig: cloud}(b) by two orders of magnitude. Thus, it is clear that this electric field has no noticeable effect on the emission picture at the discharge axis. This is visible in figure~\ref{fig: ems}(b), where we demonstrate the emission pattern estimated for the solution presented in figure~\ref{fig: cloud}(c). The emission pattern in figure~\ref{fig: ems}(b) correlates with the experimental results at moderate discharge powers (the case of a dim void). Thus, it can be hypothesized that the dim void observed in the present work corresponds to the microparticle configuration shown in figure~\ref{fig: cloud}(c). At the same time, inclusion of the radial diffusion for the conditions of the bright void does not lead to qualitative changes.

The novel void formation mechanism demonstrated above can be physically explained as follows. The gradient of the ion density in the cloud is determined by the distribution of the microparticle number density (this is also true in the radial direction). The ion flux in the radial direction can be relatively low since $G_{ion} \approx G_{d}$ inside the microparticle suspension. This is not true in the microparticle free region, where the radial gradient of $n_{i}$ is inevitably connected
with the formation of the radial ion flux. Therefore, it can be expected that the solution shown in figure~\ref{fig: cloud}(c) is realistic, at least near the discharge axis. It is clear, however, that this hypothesis can only  be confirmed (or rejected) by rigorous two-dimensional modelling studies
or by direct local measurements of electric fields and ion velocity distributions.

\section{Conclusion}

We have experimentally shown that the void (microparticle-free area) in the center of a capacitively-coupled rf complex plasma
has two regimes, which we call ``dim'' and ``bright''. 
The dim void forms at relatively low discharge powers and exhibits no emission feature associated with it.
With the increase of the discharge power, the bright emission from the void appears, 
i.e. the void experiences a transition from dim to bright regime. 
With further discharge power increase, the plasma emission concentrates
near the void boundaries, especially at higher gas pressure.

The transition between the regimes has a discontinuous character.
The threshold is manifested by a kink in the void size power dependencies.
Also, the microparticle density peak at the void boundary disappears with the power growth.

RF-period-resolved optical emission spectroscopy revealed that
the presence of the microparticles increases the emission in the entire axial extension of the discharge every half of the period in the case of the dim void, 
whereas for the bright void, the emission increase appears only between the expanding sheath and the opposite void boundary.

The bright void is reproduced by the time-averaged 1D fluid model.
The bright emission in the void is caused by the strong time-averaged electric field at the void boundary.
In this case, the void boundary stabilizes as a result of the balance of ion drag and electrostatic forces. 
This balance can only be satisfied at the elevated discharge power.

On the contrary, the dim void could not be obtained within the framework of the same 1D model: 
the bright void was closing with the decrease of the ionization rate 
as soon as the ion drag force was unable to balance the electrostatic force. 
Dim void could be, however, obtained by artificially including the radial diffusion term into the ion flux continuity equation.
Assuming radial diffusion term to be comparable to the ionisation and microparticle sink terms 
in a certain range of axial positions around the discharge midplane allowed to obtain a microparticle-free region,
whose boundary is stabilized by the week electrostatic forces. 
Electric field at the void boundary was in this case two orders of magnitude lower than that in the bright void regime. 
Therefore, it did not lead to the appearance of the emission on the void boundary. 
The modified model was able to qualitatively explain the experimentally  observed dim void formation
as well as the respective experimentally measured RF-period-resolved spatiotemporal emission profile.
At the same time, inclusion of the radial diffusion for the conditions of the bright void did not lead to qualitative changes. 

Such an approach could qualitatively reproduce the void formation and the RF-period-resolved emission patterns in case of dim and bright void.
We could therefore, demonstrate that the void in dim and bright void regimes forms due to two different mechanisms. 
Complete understanding of the dark void formation requires a self-consistent 2D model
and local measurements of the electric field or ion velocity distribution.

\appendix
\section{Characterization of the microparticle-plasma interaction}
\label{sec: DustPlasm}
The sink term due to absorption of ions by the microparticles is given by
\begin{equation}
G_{d}=n_{d}J_{i},
\end{equation}
where $J_{i}$ is the ion current collected by a single microparticle. The ion current is computed using the model presented in \cite{khrapak2005particle}:
\begin{equation}
J_{i}=J_{\rm OML}+\left[J_{\rm coll}-J_{\rm OML} \right](R_{0}/l_{ia}),
\end{equation}
where $J_{\rm OML}$ is the current derived from the orbital-motion-limited (OML) theory \cite{khrapak2009basic}, $J_{coll}$ is the current related to the ion-atom collisions, $l_{ia}$ is the mean free path for the ion-atom collisions, $R_{0}$ is the characteristic length scale of the perturbed plasma region around the microparticle. Here, $J_{\rm OML}=\sqrt{8 \pi}a_{d}^{2}n_{i}\upsilon_{i}(1-e\varphi_{d}/k_{\rm B}T_{g})$ and $J_{coll}=\sqrt{8 \pi}R_{0}^{2}n_{i}\upsilon_{i}$, where $\upsilon_{i}=\sqrt{k_{\rm B}T_{g}/m_{i}}$. The mean free path $l_{ia}$ is estimated as $l_{ia}=1/(n_{g}\bar{\sigma})$, where $\bar{\sigma}=1.32\times10^{-18}\,$m$^{2}$ is the characteristic cross-section at the collision energy of $k_{\rm B}T_{g}$ (the cross-sections are taken from \cite{phelps1994application}). The ions are assumed to interact with the microparticle via the Yukawa potential with the screening length $\lambda_{s}$. In this case, the ratio $R_{0}/\lambda_{s}$ is given by the root of the nonlinear equation $\beta \exp(-\xi)=\xi$, where $\beta=(e |\varphi_{d}|/k_{\rm B}T_{g})(a_{d}/\lambda_{s})$.\\ \indent
The ion drag force is calculated using the well-known expression \cite{fortov2005complex}:
\begin{equation}
\label{Fid}
F_{d}=\alpha \, m_{i}n_{i}\int \upsilon_{x} \upsilon \sigma(\upsilon) f(\vec{\upsilon}) d \vec{\upsilon},
\end{equation}
where $\vec{\upsilon}$ is the ion molecular velocity ($\upsilon=|\vec{\upsilon}|$), $f(\vec{\upsilon})$ is the ion velocity distribution function, $\sigma$ is the momentum transfer cross section for the ion-dust collisions, $\alpha$ is the correction factor accounting for the effect of ion-atom collisions. The ion distribution function is given by the distribution for the drift flow with constant electric field \cite{patacchini2008fully,hutchinson2013collisional}.
The cross section $\sigma$ is taken as the momentum transfer cross section for the attractive Yukawa potential \cite{khrapak2014accurate}. 
The correction factor $\alpha$ was estimated using the approximate expression presented in \cite{hutchinson2013collisional} (namely, equation~(33) in \cite{hutchinson2013collisional}). For the conditions of our study we get $\alpha \approx 1.4$ (this coefficient depends weakly on the ion velocity).
Using equation~(\ref{Fid}), the ion drag force was calculated as a function of $u_{i}$ and coupling parameter $\beta$.\\ \indent
The effective screening length for the Yukawa potential is defined using the following approximate expression \cite{khrapak2005hybrid}:
\begin{equation}
\lambda_{s}^{-2}=\lambda_{{\rm D}e}^{-2}+\tilde{\lambda}_{s}^{-2}\left[1+(u_{i}/\upsilon_{i})^{2}\right]^{-1},
\end{equation}
where $\lambda_{{\rm D}e}=\sqrt{e^{2}n_{e}/\varepsilon_{0}k_{\rm B}T_{e}}$ is the electron Debye radius and $\tilde{\lambda}_{s}$ is the screening length for the isotropic case ($u_{i} \rightarrow 0$). The screening length for the isotropic case is defined as \cite{semenov2015approximate}: 
$\tilde{\lambda}_{s}=\lambda_{{\rm D}i}\sqrt{1+0.48\sqrt{\beta_{i}}}$,
where $\beta_{i}=(e |\varphi_{d}|/k_{\rm B}T_{g})(a_{d}/\lambda_{{\rm D}i})$ and $\lambda_{{\rm D}i}=\sqrt{e^{2}n_{i}/\varepsilon_{0}k_{\rm B}T_{g}}$ is the ion Debye radius.\\ \indent
The thermophoretic force is evaluated using the conventional expression \cite{fortov2005complex}:
\begin{equation}
F_{th}=-\frac{8\sqrt{2\pi}}{15} \frac{a_{d}^{2}}{\upsilon_{a}}\lambda_{a}\frac{\partial T_{a}}{\partial x},
\end{equation}
where $T_{a}$ is the temperature of the gas atoms, $\lambda_{a}$ is the thermal conductivity and $\upsilon_{a}=\sqrt{k_{\rm B}T_{g}/m_{a}}$, with $m_{a}$ being the atom mass. Here $T_{a}$ is assumed to be close to $T_{g}$.  The thermal conductivity of argon is taken as 17.8 mW\,m$^{-1}$\,K$^{-1}$  \cite{lemmon2004viscosity}.\\\indent
The distribution of $T_{a}$ is expressed as $T_{a}=\bar{T}_{a}+\tilde{T}_{a}$, where $\bar{T}_{a}$ is the linear temperature distribution between the electrodes and $\tilde{T}_{a}$ is the solution of the heat conduction equation  in the interelectrode
region with zero boundary conditions. We assume that $F_{th}$ corresponding to the linear term $\bar{T}_{a}$ is balanced by the gravity force. Thus, the nontrivial contribution of $F_{th}$ to the force balance given by equation~(\ref{EqNd}) is related to the second term $\tilde{T}_{a}$.
The distribution of $\tilde{T}_{a}$ is found by solving the heat conduction equation:
\begin{equation}
\label{EqTa}
\lambda_{a}\frac{d^{2} \tilde{T}_{a}}{d x^{2}}=Q_{d},
\end{equation}
where $d\tilde{T}_{a}/dx=0$ at $x=0$ (the symmetry condition), $\tilde{T}_{a}=0$ at $x=L$, and $Q_{d}$ is the heat source due to the heat exchange between the gas and the dust particles. As it was discussed in \cite{pikalev2019measurement}, the effect of $Q_{d}$ is not negligible for the conditions of our study (moreover, $Q_{d}$ dominates over the contribution from the heat exchange between atoms and electrons). In this work, we estimate $Q_{d}$ as follows \cite{liu2006heat}:
\begin{equation}
Q_{d}=n_{d}\pi a^{2} \sqrt{\frac{8}{\pi}}\frac{p_{g}\upsilon_{a}}{2}
\frac{\gamma+1}{\gamma-1}\left(\frac{T_{p}}{T_{g}}-1\right),
\end{equation}
where $T_{p}$ is the microparticle surface temperature and $\gamma=5/3$. The particle temperature can be estimated by considering the balance of heat fluxes at the particle surface \cite{swinkels2000microcalorimetry,khrapak2006grain}. For the conditions of our study, this estimate gives $(T_{p}-T_{g})/T_{g} \approx 0.2 - 0.5$. For simplicity, we used the fixed value $(T_{p}-T_{g})/T_{g}=0.2$ in all computations. Equation~(\ref{EqTa}) was solved numerically using the finite difference method.

In the case without the void, the microparticle heating induces $F_{th} \sim 10^{-15}\,$N at $x=2\,$mm.
Elastic collisions with the electrons induce the thermophoretic force one order of magnitude smaller.

\ack
We would like to thank Dr. S. Khrapak for helpful discussions and 
Dr. C. Knapek for careful reading of our manuscript and valuable suggestions.

The PK-3 Plus chamber was funded by the space agency of DLR 
with funds from the federal ministry for economy and technology according to a resolution of the Deutscher Bundestag under grants No. 50WP0203, 50WM1203.
A.~Pikalev acknowledges the financial support of Deutscher Akademischer Austauschdienst (DAAD) with funds from 
Deutsches Zentrum f\"{u}r Luft- und Raumfahrt e.V. (DLR).

\bigskip
\bibliographystyle{iopart-num}

\bibliography{pikalev_void}

\providecommand{\newblock}{}
\begin{thebibliography}{10}
\expandafter\ifx\csname url\endcsname\relax
  \def\url#1{{\tt #1}}\fi
\expandafter\ifx\csname urlprefix\endcsname\relax\def\urlprefix{URL }\fi
\providecommand{\eprint}[2][]{\url{#2}}

\bibitem{fortov2005complex}
Fortov V, Ivlev A, Khrapak S, Khrapak A and Morfill G 2005 {\em Phys. Rep.\/}
  {\bf 421} 1--103

\bibitem{Morfill-RevModPhys2009}
Morfill G~E and Ivlev A~V 2009 {\em Rev. Mod. Phys.\/} {\bf 81}(4) 1353--1404

\bibitem{Ivlev2012}
Ivlev A, Löwen H, Morfill G and Royall C~P 2012 {\em Complex Plasmas and
  Colloidal Dispersions: Particle-Resolved Studies of Classical Liquids and
  Solids\/} ({\em Series in Soft Condensed Matter\/} vol~5) (World Scientific
  Publishing Co. Pte. Ltd.)

\bibitem{Nefedov-NewJPhys2003}
Nefedov A~P, Morfill G~E, Fortov V~E, Thomas H~M, Rothermel H, Hagl T, Ivlev
  A~V, Zuzic M, Klumov B~A, Lipaev A~M, Molotkov V~I, Petrov O~F, Gidzenko Y~P,
  Krikalev S~K, Shepherd W, Ivanov A~I, Roth M, Binnenbruck H, Goree J~A and
  Semenov Y~P 2003 {\em New Journal of Physics\/} {\bf 5} 33--33

\bibitem{Thomas-NewJPhys2008}
Thomas H~M, Morfill G~E, Fortov V~E, Ivlev A~V, Molotkov V~I, Lipaev A~M, Hagl
  T, Rothermel H, Khrapak S~A, Suetterlin R~K, Rubin-Zuzic M, Petrov O~F,
  Tokarev V~I and Krikalev S~K 2008 {\em New J. Phys.\/} {\bf 10} 033036

\bibitem{Pustylnik-RevSciInstrum2016}
Pustylnik M~Y, Fink M~A, Nosenko V, Antonova T, Hagl T, Thomas H~M, Zobnin A~V,
  Lipaev A~M, Usachev A~D, Molotkov V~I, Petrov O~F, Fortov V~E, Rau C,
  Deysenroth C, Albrecht S, Kretschmer M, Thoma M~H, Morfill G~E, Seurig R,
  Stettner A, Alyamovskaya V~A, Orr A, Kufner E, Lavrenko E~G, Padalka G~I,
  Serova E~O, Samokutyayev A~M and Christoforetti S 2016 {\em Review of
  Scientific Instruments\/} {\bf 87} 093505

\bibitem{Land-NewJPhys2008}
Land V and Goedheer W~J 2008 {\em New Journal of Physics\/} {\bf 10} 123028

\bibitem{Schmidt-AIPPhysPlas2011}
Schmidt C, Arp O and Piel A 2011 {\em Physics of Plasmas\/} {\bf 18} 013704

\bibitem{Stefanovic-PlasmaSrcSciTech2017}
Stefanovi{\'c} I, Sadeghi N, Winter J and Sikimic B 2017 {\em Plasma Sources
  Science and Technology\/} {\bf 26} 065014

\bibitem{Bouchoule-PlasmaSrcSciTech1993}
Bouchoule A and Boufendi L 1993 {\em Plasma Sources Science and Technology\/}
  {\bf 2} 204

\bibitem{Bouchoule-PlasmaSrcSciTech1994}
Bouchoule A and Boufendi L 1994 {\em Plasma Sources Science and Technology\/}
  {\bf 3} 292--301

\bibitem{Tachibana-PlasmaSrcSciTech1994}
Tachibana K, Hayashi Y, Okuno T and Tatsuta T 1994 {\em Plasma Sources Science
  and Technology\/} {\bf 3} 314

\bibitem{Fridman-JApplPhys1996}
Fridman A~A, Boufendi L, Hbid T, Potapkin B~V and Bouchoule A 1996 {\em Journal
  of Applied Physics\/} {\bf 79} 1303--1314

\bibitem{Mitic-NewJPhys2009}
Mitic S, Pustylnik M~Y and Morfill G~E 2009 {\em New Journal of Physics\/} {\bf
  11} 083020

\bibitem{Killer-AIPPhysPlas2013}
Killer C, Bandelow G, Matyash K, Schneider R and Melzer A 2013 {\em Phys.
  Plasmas\/} {\bf 20} 083704

\bibitem{Pustylnik-PhysRevE2017}
Pustylnik M~Y, Semenov I~L, Z\"ahringer E and Thomas H~M 2017 {\em Phys. Rev.
  E\/} {\bf 96}(3) 033203

\bibitem{Lipaev-PhysRevLett2007}
Lipaev A~M, Khrapak S~A, Molotkov V~I, Morfill G~E, Fortov V~E, Ivlev A~V,
  Thomas H~M, Khrapak A~G, Naumkin V~N, Ivanov A~I, Tretschev S~E and Padalka
  G~I 2007 {\em Physical Review Letters\/} {\bf 98}(26) 265006

\bibitem{Annaratone-PhysRevE2002}
Annaratone B~M, Khrapak S~A, Bryant P, Morfill G~E, Rothermel H, Thomas H~M,
  Zuzic M, Fortov V~E, Molotkov V~I, Nefedov A~P, Krikalev S and Semenov Y~P
  2002 {\em Physical Review E\/} {\bf 66}(5) 056411

\bibitem{Goree-PhysRevE1999}
Goree J, Morfill G~E, Tsytovich V~N and Vladimirov S~V 1999 {\em Phys. Rev.
  E\/} {\bf 59}(6) 7055--7067

\bibitem{Tsytovich-PhysRevE2001}
Tsytovich V~N, Vladimirov S~V, Morfill G~E and Goree J 2001 {\em Physical
  Review E\/} {\bf 63}(5) 056609

\bibitem{Tsytovich-PhysRevE2004}
Tsytovich V~N, Vladimirov S~V and Morfill G~E 2004 {\em Physical Review E\/}
  {\bf 70}(6) 066408

\bibitem{Vladimirov-AIPPhysPlas2005}
Vladimirov S~V, Tsytovich V~N and Morfill G~E 2005 {\em Physics of Plasmas\/}
  {\bf 12} 052117

\bibitem{Gozadinos-NewJPhys2003}
Gozadinos G, Ivlev A~V and Boeuf J~P 2003 {\em New Journal of Physics\/} {\bf
  5} 32

\bibitem{Akdim-PhysRevE2003}
Akdim M~R and Goedheer W~J 2003 {\em Physical Review E\/} {\bf 67}(6) 066407

\bibitem{Land-NewJPhys2007}
Land V and Goedheer W~J 2007 {\em New Journal of Physics\/} {\bf 9} 246

\bibitem{Goedheer-PlasmaPhysControlFusion2008}
Goedheer W~J and Land V 2008 {\em Plasma Physics and Controlled Fusion\/} {\bf
  50} 124022

\bibitem{Goedheer-JPhysD2009}
Goedheer W~J, Land V and Venema J 2009 {\em Journal of Physics D: Applied
  Physics\/} {\bf 42} 194015

\bibitem{Goedheer-ContribPlasPhys2009}
Goedheer W~J, Land V and Venema J 2009 {\em Contributions to Plasma Physics\/}
  {\bf 49} 199--214

\bibitem{Tawidian-EPS2013}
Tawidian H, Diop F, Lecas T, Gibert T and Mikikian M 2013 Void behavior and
  profile using laser induced fluorescence {\em 40th EPS Conference on Plasma
  Physics\/} (Espoo, Finland) p P1.304

\bibitem{Samsonov-PhysRevE1999}
Samsonov D and Goree J 1999 {\em Physical Review E\/} {\bf 59}(1) 1047--1058

\bibitem{Schulze-PlasmaSrcSciTehnol2006}
Schulze M, von Keudell A and Awakowicz P 2006 {\em Plasma Sources Science and
  Technology\/} {\bf 15} 556--563

\bibitem{Lagrange-JApplPhys2015}
Lagrange J~F, G{\'e}raud-Grenier I, Faubert F and Massereau-Guilbaud V 2015
  {\em Journal of Applied Physics\/} {\bf 118} 163302

\bibitem{Mikikian-NewJPhys2007}
Mikikian M, Couëdel L, Cavarroc M, Tessier Y and Boufendi L 2007 {\em New
  Journal of Physics\/} {\bf 9} 268

\bibitem{Pustylnik-AIPPhysPlas2012}
Pustylnik M~Y, Ivlev A~V, Sadeghi N, Heidemann R, Mitic S, Thomas H~M and
  Morfill G~E 2012 {\em Physics of Plasmas\/} {\bf 19} 103701

\bibitem{Rothermel-PhysRevLett2002}
Rothermel H, Hagl T, Morfill G~E, Thoma M~H and Thomas H~M 2002 {\em Physical
  Review Letters\/} {\bf 89}(17) 175001

\bibitem{Wiese-PhysRevA1989}
Wiese W~L, Brault J~W, Danzmann K, Helbig V and Kock M 1989 {\em Physical
  Review A\/} {\bf 39}(5) 2461--2471

\bibitem{Sarkar-PlasmaSrcSciTech2015}
Sarkar S, Mondal M, Bose M and Mukherjee S 2015 {\em Plasma Sources Science and
  Technology\/} {\bf 24} 035007

\bibitem{brinkmann2011plasma}
Brinkmann R 2011 {\em J. Phys. D: Appl. Phys.\/} {\bf 44} 042002

\bibitem{semenov2017moment}
Semenov I 2017 {\em Phys. Rev. E\/} {\bf 95} 043208

\bibitem{khrapak2014ion}
Khrapak S, Khrapak A, Ivlev A and Thomas H 2014 {\em Phys. Plasmas\/} {\bf 21}
  123705

\bibitem{khrapak2014simple}
Khrapak S, Khrapak A, Ivlev A and Morfill G 2014 {\em Phys. Rev. E\/} {\bf 89}
  023102

\bibitem{khrapak2015practical}
Khrapak S~A and Thomas H~M 2015 {\em Phys. Rev. E\/} {\bf 91} 023108

\bibitem{hagelaar2005solving}
Hagelaar G and Pitchford L 2005 {\em Plasma Sources Sci. Technol.\/} {\bf 14}
  722

\bibitem{biagi}
Biagi database, private communication, www.lxcat.net, retrieved on July 15,
  2020

\bibitem{sukhinin2013dust}
Sukhinin G, Fedoseev A, Antipov S, Petrov O and Fortov V 2013 {\em Phys. Rev.
  E\/} {\bf 87} 013101

\bibitem{khrapak2005particle}
Khrapak S, Ratynskaia S~V, Zobnin A, Usachev A, Yaroshenko V, Thoma M,
  Kretschmer M, H{\"o}fner H, Morfill G, Petrov O {\em et~al.\/} 2005 {\em
  Phys. Rev. E\/} {\bf 72} 016406

\bibitem{khrapak2009basic}
Khrapak S and Morfill G 2009 {\em Contrib. Plasma Phys\/} {\bf 49} 148--68

\bibitem{phelps1994application}
Phelps A~V 1994 {\em J. Appl. Phys.\/} {\bf 76} 747--53

\bibitem{patacchini2008fully}
Patacchini L and Hutchinson I~H 2008 {\em Phys. Rev. Lett.\/} {\bf 101} 025001

\bibitem{hutchinson2013collisional}
Hutchinson I and Haakonsen C 2013 {\em Phys. Plasmas\/} {\bf 20} 083701

\bibitem{khrapak2014accurate}
Khrapak S 2014 {\em Phys. Plasmas\/} {\bf 21} 044506

\bibitem{khrapak2005hybrid}
Khrapak S, Ivlev A, Zhdanov S and Morfill G 2005 {\em Phys. Plasmas\/} {\bf 12}
  042308

\bibitem{semenov2015approximate}
Semenov I, Khrapak S and Thomas H 2015 {\em Phys. Plasmas\/} {\bf 22} 053704

\bibitem{lemmon2004viscosity}
Lemmon E~W and Jacobsen R~T 2004 {\em Int. J. Thermophys\/} {\bf 25} 21--69

\bibitem{pikalev2019measurement}
Pikalev A, Pustylnik M, R{\"a}th C and Thomas H 2019 {\em J. Phys. D: Appl.
  Phys.\/} {\bf 53} 075203

\bibitem{liu2006heat}
Liu F, Daun K, Snelling D~R and Smallwood G~J 2006 {\em Appl. Phys. B\/} {\bf
  83} 355--82

\bibitem{swinkels2000microcalorimetry}
Swinkels G, Kersten H, Deutsch H and Kroesen G 2000 {\em J. Appl. Phys.\/} {\bf
  88} 1747--55

\bibitem{khrapak2006grain}
Khrapak S and Morfill G 2006 {\em Phys. Plasmas\/} {\bf 13} 104506

\end{thebibliography}

\end{document}